\documentclass[12pt,notoc]{JHEP3}

\usepackage{amsmath,amssymb,euscript,array,cite}

\def\N{{\cal N}}

\def\Z{{\bf Z}}

\def\R{\boldsymbol{R}}
\def\S{\boldsymbol{S}}
\def\Z{\boldsymbol{Z}}
\def\T{\boldsymbol{T}}

\def\Dbarslash{\,\,{\raise.15ex\hbox{/}\mkern-12mu {\bar D}}}
\def\Dslash{\,\,{\raise.15ex\hbox{/}\mkern-12mu D}}
\def\delslash{\,\,{\raise.15ex\hbox{/}\mkern-9mu \partial}}
\def\delbarslash{\,\,{\raise.15ex\hbox{/}\mkern-9mu {\bar\partial}}}

\newcommand{\EQ}[1]{\begin{equation} #1 \end{equation}}

\title{Phase Transitions of 
Orientifold Gauge Theories at Large N in Finite Volume}
\author{Timothy J. Hollowood and Asad Naqvi \\
Department of Physics,\\ University of Wales Swansea,\\
Swansea, SA2 8PP, UK.\\
E-mail: {\tt t.hollowood,a.naqvi@swan.ac.uk}}
\preprint{SWAT/06/}
\abstract{In this paper we consider the phase structure of ``orientifold''
gauge theories---obtained from unitary supersymmetric gauge theories
by replacing adjoint Majorana fermions by Dirac fermions in the
symmetric or anti-symmetric representations---in finite volume
$\S^3\times\S^1$. If the radius of the $\S^3$ is small the
calculations can be performed at weak coupling for any value of the
$\S^1$ radius. We demonstrate that there is a confinement/de-confining
type of phase transition even when the fermions have periodic
(non-thermal) boundary conditions around $\S^1$. At small radius of $\S^1$, the
theory is in a phase where charge conjugation and large non-periodic
gauge transformation are spontaneously broken. But for large radius of $\S^1$
the phase preseves these symmetries just as in the related
supersymmetric theory.}

\begin{document}


``Orientifold'' gauge theories have been the subject of much interest
due to the possibility of large $N$ equivalence
\cite{Kachru:1998ys,Strassler:2001fs}. The sobriquet
``orientifold'' comes from string theory, but field theorists may
think of it  as describing a
gauge theory which has the same field content as a
supersymmetric gauge theory but
where one replaces the adjoint Majorana fermions
by Dirac fermions transforming in either the anti-symmetric or symmetric
representation.\footnote{We are assuming a $U(N)$ or $SU(N)$ gauge
  group.} Since the dimensions of the anti-symmetric and
symmetric representations are $\tfrac12N(N\pm1)$ and that of the adjoint
is $N^2$, both theories have the same number of fermionic
degrees-of-freedom in the large $N$ limit.
The idea is that certain observables of these theories
will be equal to those of the original supersymmetric 
theory in the large $N$ limit
\cite{Armoni:2003gp,Armoni:2003fb,Armoni:2004uu,Armoni:2003yv}. This is an
intriguing possibility, but it is not easy to test the hypothesis because
in $\R^4$ these theories are strongly coupled confining gauge theories.
However, there are ways of rendering a strongly coupled theory
weakly coupled and in the weakly coupled regime can can attempt a test
of the hypothesis. The first such proposal \cite{Barbon:2005zj}
considered the theory on $\T^3\times\R$ and the conclusion was that
for small tori needed to ensure weak coupling, large $N$ equivalence
was not manifest. A different kind of philosophy
was proposed in \cite{Unsal:2006pj} who 
considered the theory on $\R^3\times\S^1$ with either periodic or
anti-periodic (thermal) boundary conditions on the fermions. 
When the size of the circle
(which we denote by $\beta$) is small, then the Wilson loop of the
gauge field around the circle 
can get a large VEV $\sim\beta^{-1}$ which breaks the gauge group
$U(N)\to U(1)^N$ at an energy scale much greater than $\Lambda_{QCD}$
and so renders the theory weakly-coupled. 
An effective action for the
eigenvalues of the Wilson loop, or Polyakov loop in the thermal case,
\EQ{
U=\text{Tr}\,\exp i\oint_{\S^1}\,A=\sum_{j=1}^Ne^{i\theta_j}\ ,
\label{wl}
}
can be computed in this regime and minimized in order to
find the ground state of the theory. The simplest kind of
orientifold gauge theory involves in addition to the gauge field, a
Dirac fermion transforming in the anti-symmetric or symmetric
representation of the gauge group. The resulting ground state
depends crucially on whether the boundary conditions on the fermions
are periodic or anti-periodic. In the periodic case, the ground state
has all the eigenvalues $\theta_i$ sitting at $\pi/2$ or all at 
$3\pi/2$ and the so there is spontaneous symmetry
breaking of the $\Z_2$ large gauge transformations which take
$\theta_i\to\theta_i+\pi$ (to be described 
more fully later) and also of charge conjugation which takes $U\to
U^*$, or $\theta_i\to-\theta_i$.\footnote{In actual fact when all the
  eigenvalues are equal, the theory is not strictly speaking weakly
  coupled since there is no Higgs mechanism.} 
For thermal boundary conditions, the ground
state is either $\theta_i=0$ or $\pi$ and so, although $\Z_2$ is also
broken, charge conjugation is preserved. 

On the contrary, in the 
related ``sister'' theory, in
this case $\N=1$ Yang-Mills, with periodic boundary conditions for the
fermions, the $\theta_i$ are always uniformly
distributed around the circle and charge conjugation is always
unbroken. This shows conclusively that large $N$
equivalence is not valid on $\R^3\times\S^1$ at small
radius with periodic boundary conditions on the fermions. 
Unfortunately, one cannot say anything about large radius (and
therefore $\R^4$) where the theory becomes strongly coupled using this
approach. The problem is that there may
be a phase transition in the orientifold theory 
for some critical radius for which $\Z_2$ and charge conjugation are
restored. Experience suggests that these kind of phase transitions do
indeed occur in the thermal case with anti-periodic boundary
conditions for the fermions. The phase transition in question is the
confinement/de-confinement transition where the de-confined phase
occurs at high temperature (small radius) and involves spontaneous
breaking of large non-periodic gauge transformations and the generation of a
condensate $\langle U\rangle\neq0$. However, with periodic boundary
conditions in the sister supersymmetric theory, this symmetry is preserved
and the confinement/de-confinement
transition does not occur. For the orientifold theory, however, we
simply do not know. In the case of thermal (anti-periodic) boundary
conditions on the fermions, \cite{Unsal:2006pj} 
shows that at high temperature (small radius) 
both the orientifold and original theories break their
group of large non-periodic gauge transformations, $\Z_2$ and $U(1)$,
respectively. The difference is that charge conjugation is preserved
in both cases and so large $N$ equivalence may be valid here.

There is a way of studying these kinds of phase transitions whilst
remaining in a weakly-coupled regime. Namely, we can investigate the
theory on $\S^3\times\S^1$. In this situation, we can keep the radius
of $\S^3$, $R\ll1/\Lambda_{QCD}$, and then study the physics as a
function of the ratio $\beta/R$ for all values of $\beta$. Of course,
one can, at the end of the day, argue that there are additional phase
transitions as $R$ varies but nevertheless the universality classes of
the transitions seen at small $R$ seem to match the expected
transitions in the strongly coupled theories on $\R^3$ at finite
temperature. Our main conclusion is that in the orientifold case,
there is a phase transition even with periodic boundary conditions.

We now turn to the calculations and, 
following the beautiful paper \cite{Aharony:2003sx} 
whose results and notion we use extensively,
we compute a Wilsonian effective action for the gauge theory on 
$\S^3\times\S^1$ to the one loop order. The only zero modes belong to the
constant mode of $A_0$, the gauge field component around
$\S^1$:
\EQ{
\alpha=\frac1{\text{Vol}\,\S^3\times\S^1}\int_{\S^3\times\S^1} A_0\ .
} 
We can use global gauge transformation to diagonalize $\alpha$:
\EQ{
\alpha=\beta^{-1}\text{diag}(\theta_i)\ .
\label{kii}
}
The $\theta_i$ are angular variables since there are large gauge
transformations (but periodic around $\S^1$)
that take $\theta_i\to\theta_i+2\pi$. Physically, the gauge
invariant quantity is the Wilson loop \eqref{wl}.
On top of this there are 
additional large gauge transformations that are only periodic on
$\S^1$ up
to an element of the centre $\Gamma$ of the gauge group a quantity
that depends on the matter content. These large gauge
  transformation form a group themselves 
isomorphic to $\Gamma$ and we will refer
  to it as $\tilde\Gamma$. If there is only adjoint
  matter and the gauge group is $SU(N)$ 
then $\tilde\Gamma=\Z_N$, while for gauge group $U(N)$ we have
$\tilde\Gamma=U(1)$. In the presence of matter the centre and hence
$\tilde\Gamma$ can be a smaller subgroup of $\Z_N$ or $U(1)$, respectively.
If there were only adjoint matter then  in the $SU(N)$ case these
non-periodic 
large gauge transformations take $\theta_i\to\theta_i+2\pi/N$ and so 
transform $U$ by an $N$-th root of unity. In the $U(N)$ case, the
transformations are $\theta_i\to\theta_i+a$ for $0\leq a<2\pi$. Hence,
 strictly speaking, the gauge invariant observables are, for example,
 $|U|$. Spontaneous symmetry breaking of this $\tilde\Gamma$
  symmetry occurs when $\langle |U|\rangle\neq0$.

The radiative corrections at the one loop level are obtained by
taking the constant mode \eqref{kii} as a background VEV and
integrating out all the massive modes of the fields. The this end, we
shift $A_0\to A_0+\alpha$ and then the one-loop
contribution involves the logarithm of the resulting functional
determinants and which depend on $\alpha$ in a non-trivial way.

The detailed calculations have been performed in \cite{Aharony:2003sx} 
and so our
discussion will be brief. Each field is expanded in terms of harmonics
on $\S^3\times\S^1$ and keeping only the quadratic terms and
integrating out the fields, a typical contribution to the effective action
is of the form
\EQ{
\pm\tfrac12\text{Tr}\log(-\tilde D_0^2-\Delta)\ ,
\label{sos}
}
the $\pm1$ being for bosons and fermions, respectively. In the
above, $\tilde D_0=\partial_0+i\alpha$ and so includes the coupling to
the VEV, and $\Delta$ is the Laplacian on
$\S^3$ appropriate to the tensorial nature of the field on $\S^3$. 
The background VEV $\alpha$ acts as a generator of the Lie
algebra of $SU(N)$ in the representation of the gauge group
appropriate to the field and the trace includes a trace over that
representation of the gauge group. The eigenvalues of $\partial_0$ are
simply $2\pi in/\beta$, $n\in\Z$, while the eigenvectors of the
Laplacian on $\S^3$ are labelled by the angular momentum $\ell$:
\EQ{
\Delta \psi_\ell=-\varepsilon_\ell^2\psi_\ell\ ,
}
and we denote their degeneracy as $d_\ell$. The data
$\varepsilon_\ell$ and $d_\ell$ depend on the 
field type as we list below.

(i) {\bf Scalars}. There are two kinds of scalar fields which have the
same set of eigenvectors. Firstly, for conformally coupled 
scalars\footnote{These have a mass term involving the Ricci scalar of
  the manifold, in this case simply $R^{-1}$.} 
we have $\varepsilon_\ell=R^{-1}(\ell+1)$. On the other hand, 
for minimally coupled scalars
$\varepsilon_\ell=R^{-1}\sqrt{\ell(\ell+2)}$. Both types have a 
degeneracy $d_\ell=(\ell+1)^2$ with $\ell\geq0$. 

(ii) {\bf Spinors}. For 2-component complex
spinors,\footnote{This representation is actually reducible on $\S^3$ into two
  real 2-component spinors but corresponds to a Majorana spinor on 
$\S^3\times \S^1$.} we have
$\varepsilon_\ell=R^{-1}(\ell+1/2)$ and
$d_\ell=2\ell(\ell+1/2)$ with $\ell>0$. 

(iii) {\bf Vectors}. Here the situation is more complicated. 
A vector field $V_i$ can be decomposed
into the image and the kernel of the covariant derivative:
$V_i=\nabla_i\chi+B_i$, with $\nabla^iB_i=0$. The eigenvectors 
for the closed part, $B_i$, have
$\varepsilon_\ell=R^{-1}(\ell+1)$ and $d_\ell=2\ell(\ell+2)$ with
$\ell>0$. 
On the
other hand, the exact part $\nabla_i\chi$ has
$\varepsilon_\ell=R^{-1}\sqrt{\ell(\ell+2)}$ with degeneracy
$d_\ell=(\ell+1)^2$ but with $\ell>0$ only.  

Notice that both spinors and vectors have no zero ($\ell=0$) modes on
$\S^3$. This is why in a pure gauge theory the only field with a zero
mode is $A_0$ which is a scalar on $\S^3$.

It is a standard calculation using the identity
$\prod_{n=1}^\infty(1+x^2/n^2)=\sinh(\pi x)/(\pi x)$
to show that \eqref{sos} is equal, up to an infinite additive constant, to
\EQ{
\sum_{\ell=0}^\infty d_\ell\Big\{\beta\varepsilon_\ell
-\sum_{n=1}^\infty\frac1n 
e^{-n\beta \varepsilon_\ell}\text{Tr}\,\cos(\beta\alpha)\Big\}\ .
\label{saa}
}
The first term here involves the Casimir energy and since it is independent
of $\alpha$ will play no r\^ole in our story and we will subsequently
drop it. 

The sums over the angular momentum in \eqref{saa} can be performed
explicitly for each of the tensor types on $\S^3$. Firstly for
conformally coupled scalars
\EQ{
z_s(x)=\sum_{\ell=0}^\infty(\ell+1)^2x^{-(\ell+1)}=
\frac{x(1+x)}{(1-x)^3}\ ,
}
whilst for spinors
\EQ{
z_f(x)=2\sum_{\ell=1}^\infty\ell(\ell+1/2)x^{-(\ell+1/2)}=\frac{4x^{3/2}}
{(1-x)^3}\ ,
}
and finally for closed vectors
\EQ{
z_v(x)=2\sum_{\ell=1}^\infty\ell(\ell+2)x^{-(\ell+1)}=\frac{x(6x-2x^2)}{(1-x)^3}\ ,
\label{jhh}
}
where $x=e^{-\beta/R}$. We will not need the sums for minimally coupled
scalars and exact vectors.

Before we consider theories with matter fields, let us first consider pure
Yang-Mills where we have to face the issue of gauge fixing. One can
follow the non-covariant gauge fixing in \cite{Aharony:2003sx},
however it is perhaps simpler to use a conventional Faddeev-Popov
procedure and choose Feynman gauge. 
The gauge field $A_\mu$ includes $A_0$ which
transforms as a minimally coupled scalar on $\S^3$, while
$A_i=B_i+\nabla_i\chi$. The ghosts transform as minimally coupled
scalars on $\S^3$ but contribute with a $-1$ in \eqref{sos} since they
are Grassmann valued. 
The $\ell>0$ contributions from $A_0$, $\nabla_i\chi$ and the
ghosts all cancel leaving only a net contribution from the $\ell=0$
modes (since exact vector do not have an $\ell=0$ mode) of the form
\EQ{
\sum_{n=1}^\infty\frac1n
\text{Tr}\cos(n\beta\alpha)=\sum_{n=1}^\infty\frac1n\sum_{ij=1}^\infty
\cos n(\theta_i-\theta_j)\ .
\label{jac}
}
This part is precisely the exponentiation of the Jacobian that
converts the integrals over the $\theta_i$ into an integral over the
unitary matrix $U=\text{diag}(e^{i\theta_i})$:
\EQ{
\int\prod_{i=1}^Nd\theta_i\,\exp\Big\{
\sum_{n=1}^\infty\frac1n\sum_{ij=1}^N\cos n(\theta_i-\theta_j)\Big\}
\propto\int\prod_{i=1}^Nd\theta_i\,\prod_{i<j}\sin^2\Big(
\frac{\theta_i-\theta_j}2\Big)
=\int dU\ .
}
However, we will leave the Jacobian in the exponent since it must be
considered as part of the effective action for the eigenvalues.

The remaining modes are the closed vectors $B_i$ and these contribute
as in \eqref{saa} with $d_\ell=2\ell(\ell+2)$ and
$\varepsilon_\ell=R^{-1}(\ell+1)$. Using the 
sum \eqref{jhh} and including the Jacobian term
in \eqref{jac}, the full effective action is simply
\EQ{
S(\theta_i)=\sum_{n=1}^\infty\frac1n\big(1-z_v(x^n)\big)
\sum_{ij=1}^N\cos n(\theta_i-\theta_j)\ .
\label{ioi}
}
The phase structure as a function of the temperature is determined by
minimizing $S(\theta_i)$. 

As the temperature changes from low to high, the parameter $x$ varies
from 0 to 1. At low temperatures, the pre-factor of the cosine is
positive and the eigenvalues effectively repel 
one another and for large $N$ form a
uniform distribution around the circle. As $x$ increases the factor
$1-z_v(x)$ changes sign at some critical temperature $T=T_c$ which can
be found by solving $z_v(e^{-1/RT_c})=1$. Beyond
this the eigenvalues attract each other and in the limit of very 
high temperatures the
distribution of eigenvalues becomes a delta function at some 
arbitrary point $\theta_0$ around the circle. Notice that the
transition is driven by the $n=1$ term in \eqref{ioi}. Of course in a situation
with $N$ finite there is no genuine symmetry breaking in finite volume
and one should integrate over the modulus $\theta_0$. However, if we
take the large $N$ limit, then a sharp phase transition does indeed 
occur at $T_c$ and the high temperature phase spontaneously breaks the
$\tilde\Gamma$ symmetry.\footnote{This is $U(1)$ in the case of pure
Yang-Mills with a $U(N)$ gauge group or if the gauge group is $SU(N)$
then $\theta_0=2\pi n/N$, $n\in\Z$, and $\tilde\Gamma=\Z_N$.} The order
parameter is the expectation value of the Wilson/Polyakov loop with $\langle
U\rangle=0$ in the low temperature phase and $\langle
U\rangle=Ne^{i\theta_0}\neq 0$ in the high temperature phase. This
pattern of symmetry breaking is precisely what one expects for the 
confinement/de-confinement phase transition in the strongly-coupled
theory on $\R^3$. Another order parameter is the 
effective action itself. At low temperatures $S=0$ while above the
transition $S={\cal O}(N^2)$ as one would expect if the colour
degrees-of-freedom where being de-confined. 

Another way to analyze the effective action in the large $N$ limit is
to represent the distribution of the eigenvalues $\theta_i$ by a 
density $\rho(\theta)$ normalized so that
$\int_0^{2\pi}d\theta\,\rho(\theta)=1$ and replace
$\sum_{i=1}^Nf(\theta_i)\to
N\int_0^{2\pi}d\theta\,\rho(\theta)f(\theta)$. 
To this end, it is useful to define the Fourier
components
\EQ{
\rho_n^+=\int_0^{2\pi}d\theta\,\rho(\theta)\cos(n\theta)\ ,\qquad
\rho_n^-=\int_0^{2\pi}d\theta\,\rho(\theta)\sin(n\theta)\ ,
}
with $\rho_0^+=1/(2\pi)$,
in terms of which the effective action is
\EQ{
S(\rho_n^\pm)=
\frac{N^2}{2\pi}\sum_{n=1}^\infty\big\{V_n^+(T)(\rho_n^+)^2+V_n^-(T)
(\rho_n^-)^2\big\}\
,
\label{fge}
}
where, in this case,
\EQ{
V_n^+(T)=V_n^-(T)=\frac{2\pi}n\big(1-z_v(x^n)\big)\ .
}
At low temperature, all the $V_n^\pm(T)$ are positive and so
$\rho_n^\pm=0$, $n>0$. This means that only $\rho_0^+=1/(2\pi)$ is
non-vanishing corresponding to the uniform distribution. 
As $x$ increases $V_1^\pm(T)$ change sign at the critical temperature
and at $T=T_c$ the first harmonics can be non-vanishing
corresponding to
\EQ{
\rho(\theta)=\frac1{2\pi}\big(1+t\cos(\theta-\theta_0)\big)
}
for arbitrary $t$ and $\theta_0$. At $T=T_c$, $0\leq t\leq 1$ 
parameterizes a flat direction. 
Note that $V_n^\pm(T)$ for $n>1$ change sign at
a higher temperature and so the transition is determined solely by
$V_1^\pm(T)$. For $T>T_c$ one can show that $\rho(\theta)$ develops
a gap at $\theta=\theta_0+\pi$ which grows as $T$ increases so that at very
high temperatures $\rho(\theta)=\delta(\theta-\theta_0)$. The order
of the transition turns out to be a rather subtle issue that depends
on higher orders in perturbation theory. 
Remarkably, the necessary three-loop calculation was performed 
for pure Yang-Mills in
\cite{Aharony:2005bq} where it was shown that the transition 
is first order. 

{\sl $\N=1$ Yang-Mills}

In this theory, there is a adjoint-valued Majorana (or Weyl) fermion
in addition to the gauge field. Now that the theory has fermions there
are two possibilities for the boundary conditions of the fermions
around the circle. In the thermal
case, the fermions have anti-periodic boundary conditions and
supersymmetry is broken while the
other possibility is to have supersymmetry
preserving periodic boundary conditions.
The effective action is
\EQ{
S(\theta_i)=\sum_{n=1}^\infty\frac1n\big(1-z_v(x^n)+\sigma_nz_f(x^n)\big)
\sum_{ij=1}^N\cos n(\theta_i-\theta_j)\ .
}
where\footnote{This can be derived by shifting the Matsubara
  frequencies $2\pi n/\beta$ by $\pi/\beta$ in the anti-periodic case.}
\EQ{
\sigma_n=\begin{cases}(-1)^n & \text{thermal (anti-periodic)}\\
1 & \text{periodic}\ .\end{cases}
}
Or one can write it as \eqref{fge} with
\EQ{
V_n^+(T)=V_n^-(T)=\frac{2\pi}n\big(1-z_v(x^n)+\sigma_nz_f(x^n)\big)\ .
}

In this case the behaviour depends crucially on whether we have 
thermal or periodic boundary conditions for the fermions. In the
thermal case, as $T$ increases $V_1^\pm(T)$ change when $z_v(x)+z_f(x)=1$
and there is a phase transition of exactly the same
kind as in the pure gauge theory. On the other hand if we choose periodic
boundary conditions then $V_n^\pm(T)$ are always positive and no
transition occurs. In this case the system is always in the state with
a uniform distribution of eigenvalues. 

{\sl The orientifold theory}

This theory has a Dirac fermion in the anti-symmetric or symmetric
representation\footnote{This is equivalent to having
one Weyl in the anti-symmetric and one in its complex conjugate
representation.} as
well as the gauge field. 
In this case, with gauge group $U(N)$, the
group of large non-periodic gauge transformations is $\tilde\Gamma=\Z_2$.

In the anti-symmetric representation, the
eigenvalues of $\alpha$ are $\theta_i+\theta_j$, $i<j$, along with
$-\theta_i-\theta_j$ , $i<j$, for its conjugate.\footnote{In the symmetric
representation the eigenvalues include the same but in addition the ones
with $i=j$. For the most part we shall consider the anti-symmetric
representation but at large $N$ the conclusions for the symmetric
representation will be identical.} The group
$\tilde\Gamma=\Z_2$ is generated by $\theta_i\to\theta_i+\pi$. 
Using all the formulae above, one finds that the
the effective action is\footnote{Note that we have
  restricted the sum to $i\neq j$ for the anti-symmetric
  representation, however, this has no effect on the terms arising
  from the gauge field which are independent of the eigenvalues when $i=j$.}
\EQ{
S(\theta_i)=\sum_{n=1}^\infty\frac1n\sum_{i\neq
j=1}^N\Big\{\big(1-z_v(x^n)\big)
\cos n(\theta_i-\theta_j)+\sigma_nz_f(x^n)\cos n(\theta_i+\theta_j)\Big\}\ ,
\label{ors}
}
Before we consider the large $N$ limit, it is quite instructive to
consider the case of $N=2$. As in other cases, changes of state
are driven solely by the $n=1$ terms which in this case are
\EQ{
S(\theta_i)=2\big(1-z_v(x)\big)
\cos(\theta_1-\theta_2)+2\sigma_1z_f(x)\cos(\theta_1+\theta_2)+\cdots\ .
}
It is a simple matter to minimize these $n=1$ terms 
with respect to $\theta_1$ and
$\theta_2$. Notice that $z_f(x)$ is always positive, whereas $1-z_v(x)$
is positive at low temperature and negative at high temperature. 
In the thermal case, $\sigma_1=-1$ and therefore the low temperature
phase has eigenvalues $(\pi/2,3\pi/2)$. There is a transition at
$z_v(x)+z_f(x)=1$ and in the high temperatures phase one has the 
two states $(0,0)$ or $(\pi,\pi)$. Of course, since we are in finite
volume we have to sum over the two high temperature states and the
$\Z_2$ symmetry is restored. However, below we shall consider the
large $N$ limit where there is a genuine sharp phase transition and symmetry
breaking. 

Now we turn to the case of periodic boundary conditions for which 
$\sigma_1=1$. In this case, at low temperature we have the unique
state $(0,\pi)$ but there is also a transition when $z_v(x)+z_f(x)=1$. The
high temperature phase has two solutions $(\pi/2,\pi/2)$ or
$(3\pi/2,3\pi/2)$. The pattern of eigenvalues matches those found for
the theory on $\R^3\times\S^1$ in \cite{Unsal:2006pj}.

While the finite $N$ case is interesting, in order to have a genuine
phase transition we need to extend the discussion to the 
large $N$ limit. In this case, we find
\EQ{
V_n^\pm(T)=\frac{2\pi}n\big(1-z_v(x^n)\pm\sigma_n z_f(x^n)\big)\ .
}
At low temperatures both $V_n^\pm(T)$ are positive for both thermal
and periodic boundary conditions. In this case, the
ground state consists of a uniform distribution of eigenvalues
$\rho(\theta)=1/(2\pi)$. As $T$ increases there is a transition at
$z_v(x)+z_f(x)=1$ where $V_1^+(T)$ changes sign, for thermal boundary
conditions. This is exactly the same critical temperature as in the
$\N=1$ supersymmetric theory.
In this case $V_1^-(T)$ remains positive. The situation
for periodic boundary conditions is the reverse of this:
$V_1^-(T)$ changes sign and $V_1^+(T)$ remains positive. Notice that
the critical temperature is the same in both cases.
At the critical temperature the distribution can
develop the $\cos\theta$ harmonic, in the thermal case, and the
$\sin\theta$ harmonic, in the periodic case. In the high temperature phase
the $\Z_2$ symmetry $\theta\to\theta+\pi$ is spontaneously broken with 
the limiting distributions $\rho(\theta)=\delta(\theta)$ or
$\delta(\theta-\pi)$, in the thermal case, and $\delta(\theta-\pi/2)$
and $\delta(\theta-3\pi/2)$, in the periodic case. Notice that in the
periodic case, charge conjugation symmetry $\rho(\theta)\to\rho(-\theta)$ 
is also spontaneously broken, but this does not
occur in the thermal case. 

In the periodic case, the two phases match precisely those found in
\cite{Unsal:2006pj} for the theory on $\R^3\times\S^1$ for small radius.
However, as the $\S^1$ in our
case de-compactifies, we find a phase transition to the uniform
distribution of eigenvalues just as in the sister supersymmetric
theory. This does not prove definitively that such a transition
will also occur in the theory on $\R^3\times\S^1$ since this is a
strongly-coupled theory and perturbative or semi-classical methods are
not valid. 

Notice that the same formalism is also applicable to the case of the
symmetric representation by simply extending the sums in \eqref{ors} to
include the $i=j$ terms. These terms are sub-leading at large $N$ and
so do not affect the conclusions.

{\sl The orientifold $\N=4$ theory}

This theory has the same matter content as the $\N=4$ theory apart from
the fact that the four adjoint Majorana fermions 
are replaced by Dirac fermions in the symmetric or anti-symmetric
representation of the gauge group \cite{Angelantonj:1999qg}. 

For the $\N=4$ theory which has, in addition to the gauge field 6
conformally coupled scalars and 4 Majorana fermions all transforming in the
adjoint representation of the gauge group, we have
\EQ{
V_n^+(T)=V_n^-(T)=\frac{2\pi}n\big(1-z_v(x^n)-6z_s(x^n)+4\sigma_n(x)z_f(x^n)
\big)\
. 
}
In the thermal case, there is a phase transition when
$z_v(x)+6z_s(x)+4z_f(x)=1$ in the same universality class as in the
pure Yang-Mills case discussed above. In the periodic case,
$1-z_v(x)-6z_s(x)+4z_f(x)$ is always positive and no phase transition occurs.

For the orientifold version, the fermions come in either the symmetric
or anti-symmetric representation, and in these cases,
\EQ{
V_n^\pm(T)=\frac{2\pi}n\big(1-z_v(x^n)-6z_s(x^n)\pm4\sigma_n(x)z_f(x^n)\big)\
.
}
Hence in both the thermal and periodic cases there is a phase transition
at $z_v(x)+6z_s(x)+4z_f(x)=1$ in the same universality classes as the
orientifold theory described above. So in the thermal case, in the
high temperature case $\Z_2$ is spontaneously broken, but not charge
conjugation, whilst in the periodic case both $\Z_2$ and charge
conjugation are spontaneously broken. 

It would be interesting to relate this weak coupling picture to the
gravity dual. For the $\N=4$ theory, Witten argued that the phase
transition at strong coupling in the thermal case, should correspond
to the Hawking-Page transition between thermal AdS (Euclidean $AdS_5$
with a periodic identification) and the Euclidean AdS big black hole
\cite{Witten:1998zw}. The big black hole describes the de-confined
phase at high temperature. Both geometries have a one cycle on the
boundary, however, in the black hole case the cycle is contractable as
one goes into the interior forming a cigar in the bulk geometry. 
In thermal AdS the one cycle
is not contractable. These features match the fact that in the high
temperature phase there is a 
non-vanishing Polyakov loop at weak coupling which in the gravity dual
arises from a string world-sheet 
that wraps the cigar. For thermal AdS, there is no 
contractable one cycle matching the fact that the Polyakov
loop vanishes in the low temperature phase. Another symptom of the
fact that the one cycle on the boundary of the black hole
is contractable, is that only 
fermions with anti-periodic boundary
conditions can be supported. Hence there is no
Hawking-Page transition for periodic
fermion boundary conditions matching perfectly with the weak coupling picture.

For the orientifold theory
the puzzle is that there is a phase transition even with periodic
boundary conditions which suggest a Hawking-Page type transition even
though the black hole geometry cannot support periodic fermions.
Clearly this issue deserves further attention.

\vspace{1cm}

We would like to thank our colleagues Prem Kumar and Adi Armoni for
useful conversations. AN is supported by a PPARC Advanced Fellowship.

\end{document}